\documentclass[modern, dvipsnames]{aastex631}
\usepackage[T1]{fontenc}

\usepackage{graphicx}
\usepackage{float}
\usepackage{xcolor}
\usepackage{transparent}
\usepackage{bm}
\usepackage{amsmath}
\usepackage{frame}
\usepackage{multirow}
\usepackage{xcolor}
\usepackage[colorinlistoftodos]{todonotes}
\usepackage{listings}
\usepackage{DejaVuSansMono} 
\usepackage{tcolorbox}
\usepackage[utf8]{inputenc}
\usepackage{fontawesome}

\newcommand{\prose}{\textsf{prose}}
\newcommand{\AIJ}{\textsf{AIJ}}
\newcommand{\AstroImageJ}{\textsf{AstroImageJ}}
\newcommand{\Block}{\textit{Block}}
\newcommand{\Image}{\textit{Image}}
\newcommand{\Telescope}{\textit{Telescope}}

\newcommand{\Blocks}{\textit{Blocks}}
\newcommand{\Sequence}{\textit{Sequence}}
\newcommand{\Sequences}{\textit{Sequences}}

\definecolor{linkcolor}{rgb}{0.1216,0.4667,0.7059}
\newcommand{\codeicon}{{\color{RoyalBlue}\faFileCodeO}}
\newcommand{\codelink}[1]{\href{}{\codeicon}\,\,}

\newcommand{\prosegithub}{\href{https://prose.readthedocs.io}{https://prose.readthedocs.io}}

\newcommand{\footlink}[1]{\footnote{\href{#1}{#1}}}

\newcommand{\aijobs}{26\,}

\lstdefinestyle{mystyle}{
    commentstyle=\color{gray!50},
    keywordstyle=\color{Bittersweet},
    stringstyle=\color{RoyalBlue},
    basicstyle=\fontsize{7.5}{11}\fontfamily{DejaVuSansMono-TLF}\selectfont,
    breakatwhitespace=false,         
    breaklines=true,
    rulecolor=\color{black!15},
    numbers=left,
    numberstyle=\fontsize{7}{11}\fontfamily{DejaVuSansMono-TLF}\selectfont\color{gray!50},
    framerule=0pt,
    breakindent=0pt,
    resetmargins=true,
    numbersep=5pt,
    frame=single
}
\tcbset{
  colframe=black!15,
  boxrule=0.5pt,
  arc=0.5mm,
  colback=gray!0
}

\newcommand{\STAR}{STAR Institute, University of Liège, Allée du 6 Août 19C, B-4000 Liège, Belgium}
\newcommand{\Astrobio}{Astrobiology Research Unit, University of Liège, Allée du 6 Août 19C, B-4000 Liège, Belgium}
\newcommand{\Bern}{Center for Space and Habitability, Gesellsschaftstrasse 6, 3012 Bern, Switzerland}

\shortauthors{Garcia et al.}

\begin{document}
\title{\prose{}: A Python framework for modular astronomical images processing}

\author[0000-0002-4296-2246]{Lionel J. Garcia}
\author{Mathilde Timmermans}
\affiliation{\Astrobio}

\author{Francisco J. Pozuelos}
\affiliation{\STAR}
\affiliation{\Astrobio}

\author{Elsa Ducrot}
\affiliation{\Astrobio}

\author{Michael Gillon}
\affiliation{\Astrobio}

\author{Laetitia Delrez}
\affiliation{\Astrobio}
\affiliation{\STAR}

\author{Robert D. Wells}
\affiliation{\Bern}

\author{Emmanuel Jehin}
\affiliation{\STAR}

\begin{abstract}
	To reduce and analyze astronomical images, astronomers can rely on a wide range of libraries providing low-level implementations of legacy algorithms. However, combining these routines into robust and functional pipelines requires a major effort which often ends up in instrument-specific and poorly maintainable tools, yielding products that suffer from a low-level of reproducibility and portability. In this context, we present \prose{}, a Python framework to build modular and maintainable image processing pipelines. Built for astronomy, it is instrument-agnostic and allows the construction of pipelines using a wide range of building blocks, pre-implemented or user-defined. With this architecture, our package provides basic tools to deal with common tasks such as automatic reduction and photometric extraction. To demonstrate its potential, we use its default photometric pipeline to process \aijobs TESS candidates follow-up observations and compare their products to the ones obtained with \AstroImageJ{}, the reference software for such endeavors. We show that \prose{} produces light curves with lower white and red noise while requiring less user interactions and offering richer functionalities for reporting.
\end{abstract}




\section*{Introduction}

When an astronomical observation is completed, one has to process its raw products into a desired output. Observatories and survey-specific pipelines are sometimes available but general or custom software are often required, offering higher flexibility and control. These tools and their development benefit from legacy and modern algorithms, implemented into community-maintained libraries (like \texttt{IRAF} \citep{iraf} or \texttt{astropy} \citep{astropy:2013,astropy:2018}). However developing a custom option requires time and effort. Indeed, once the algorithms are selected, their implementations must be assembled into a code ingesting raw FITS images. To reach an optimized output, a fine-tuning of this code is often required, from its individual processing components up to the complete pipeline, including complex I/O operations and architecture choices. In its final version, the software is generally frozen and maintained when it breaks. These tools, whether custom or third-party, seldom provide a complete analysis framework such that their end products are exploited into scripting languages such as Julia\footlink{https://julialang.org/} or Python\footlink{https://www.python.org/}.
\bigskip\\
We propose an alternative for that. Pipelines can generally be described as a sequence of individual data-processing units. These processing units (hereafter called \Blocks) are fundamentally common from one pipeline to another. They complete independent tasks using algorithms widely used by the whole community. Hence, in a framework where \Blocks{} are all inter-compatible, one would only have to focus on which algorithms to use and assemble \Blocks{} into a modular pipeline, with no need for software architecture considerations.
\bigskip\\
In this paper, we present \prose{}\footlink{https://prose.readthedocs.io}, a Python package acting as a framework to build modular pipelines for astronomy, allowing rapid iterations toward specific needs and rich scientific products. Its architecture allows the creation of inter-compatible \Blocks{}, many being pre-implemented and available in the initial package. Developed in the context of transiting exoplanet observations, it contains tools specific to high-precision point-source photometry. However, the aim of \prose{} is to be versatile and to adapt to a variety of astronomical datasets. 
\bigskip\\
In \autoref{prose}, we present the key concepts behind \prose{}, as well as some of the pre-implemented blocks it features. In \autoref{pipeline}, we present a pipeline, developed with \prose{}, built to reduce raw astronomical images and perform high-precision differential photometry. To remain fast on domestic computers, this pipeline relies on two algorithms: a fast registration technique based on \citet{Lang_2010} and a neural network based centroiding method, both described in details in \autoref{pipeline}. In x\autoref{aij}, we assess our built-in photometric pipeline's performances by processing \aijobs TESS candidates follow-up observations, mainly from the TRAPPIST-South telescope \citep{Trappist, trapmg}, and compare their final light curves to those obtained with \AstroImageJ{} (\AIJ{}, \citealt{Collins2017}), the main software used by the TESS Follow-up Observing Program\footlink{https://tess.mit.edu/followup/} working group (TFOP) to process ground-based transit follow-up observations.

\newpage
\section{\prose{}} \label{prose}

\subsection{Images, Blocks and Sequences}

A \textit{pipeline}, strictly speaking, is a series of connected tubes running a fluid. In the scientific literature, the word refers to processing pipelines in which data are flowing, going through processing units as in tubes. Image processing algorithms are implemented as standalone codes which are easy to run on a single image. However, assembling multiple low-level codes, successively ran on a set of images requires a complete architecture to be built and tested. The final result is often monolithic, far from where we take the word \textit{pipeline} from. Of course, the flexibility offered by lower-level codes is essential but we believe generalizing the approach of building pipelines for astronomical image processing comes with great advantages.
\bigskip\\
\prose{} contains the structure for such a modular concept to work by providing three key objects: \textit{Images}, \textit{Blocks} and \textit{Sequences}. An \textit{Image} encapsulates the data and metadata of an exposure frame. A \textit{Block} is a single processing unit that acts on an \textit{Image}. Finally, a \Sequence{} is a series of \Blocks{}. A pipeline is then an assembly of \textit{Sequences}.

\begin{figure}[H]
	\centering
	\includegraphics[width=0.7\linewidth]{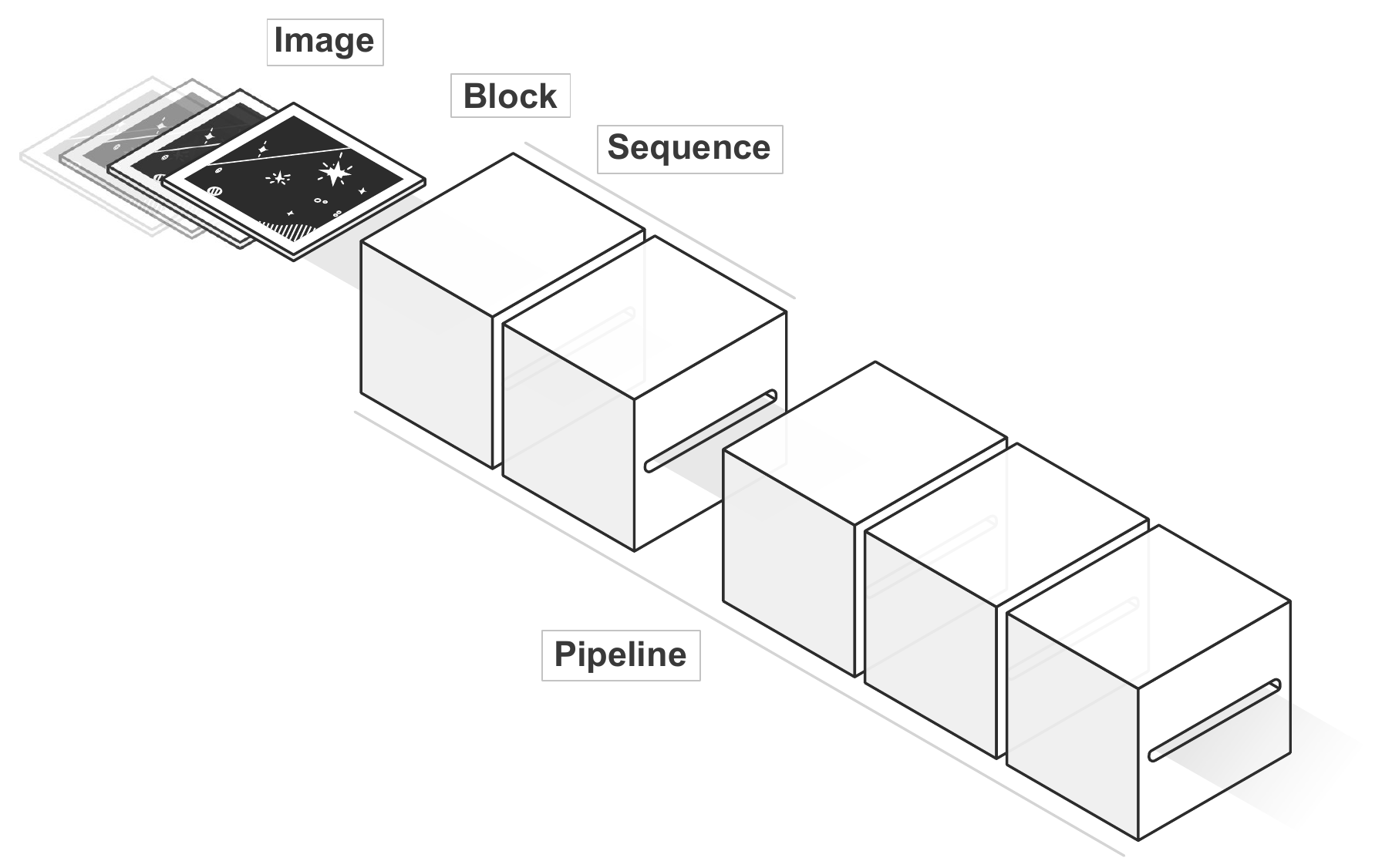}
	\caption{Illustration of \prose{} key objects}
  \end{figure}

  \subsection{Blocks}\label{blocks}

With this architecture, one can focus on designing robust and reusable \Blocks{}, well tested and independently maintainable. To do that, \prose{} comes with pre-implemented \Blocks{} featuring common state-of-the-art algorithms, some being listed in \autoref{tab:blocks}. In this table, the symbol * refers to \Blocks{} using implementations from other sources, like \texttt{PhotutilsAperturePhotometry} which relies on the aperture photometry implementation of \texttt{photutils}\footnote{\href{https://photutils.readthedocs.io}{https://photutils.readthedocs.io}}, or \texttt{SEDetection} using SExtractor \citep{sextractor} through the \texttt{sep}\footlink{https://sep.readthedocs.io} Python package.\bigskip\\

\begin{table}
  \centering
  \begin{tabular}{lll}
    \textit{Calibration and utils} & \textit{Detection} & \textit{PSF estimate} \\
	\hline
    \vspace{-0.2cm}\\
    \texttt{Calibration} & \texttt{DAOFindStars}*    & \texttt{FastGaussian} \\
    \texttt{Flip}        & \texttt{IrafFindStars}*   & \texttt{Gaussian2D}\\
    \texttt{Trim}        & \texttt{SEDetection}*     & \texttt{Moffat2D} \\
    \texttt{MedianStack} & \texttt{SegementedPeaks}  &  \\
    \texttt{Video}       &                           &  \\
    \\
    \textit{Alignment and registration}  & \multicolumn{2}{l}{\textit{Photometry}} \\
	\hline
    \vspace{-0.2cm}\\
    \texttt{XYShift} & \multicolumn{2}{l}{\texttt{PhotutilsAperturePhotometry}*} \\
    \texttt{Twirl} & \multicolumn{2}{l}{\texttt{SEAperturePhotometry}*} \\
    \texttt{AffineTransform} &  & \\
    \texttt{BalletCentroid} &  & \\
    \\
  \end{tabular}
  \caption{List of some base \Blocks{} currently included in \prose{}. * refers to \Blocks{} whose implementation is mainly coming from other packages or libraries.}
  \label{tab:blocks}
\end{table}

Describing all base \Blocks{} of \prose{} is beyond the scope of this paper, and we refer the interested reader to the \prose{} online documentation\footlink{https://prose.readthedocs.io}.

\subsection{Telescope}
\Blocks{} are likely to use \Image{} characteristics to complete their tasks. In the case of FITS images, this information is located in their headers, whose definition may differ between observatories. To deal with this variety, \prose{} provides a \Telescope{} object containing a semantic dictionary of keywords used to retrieve physical quantities from image headers. Once a new \Telescope{} is declared, its dictionary is saved and used whenever its name appears, filling the appropriate \Image{} metadata. This mechanism, also found in general reduction tools like \AIJ{}, allows \prose{} to be instrument-agnostic but is only suited to single extension FITS products containing simple headers. Images from the Vatican Advanced Technology Telescope 4K camera (VATT4k), for example, are taken by two amplifiers, leading to FITS images spread over two extensions. Each of these extensions contains data as well as useful header information, while more general characteristics can be found in a separate primary header. To allow such images to be combined and processed, \prose{} allows custom \Image{} loaders to be developed (which in the case of the VATT4k images, for example, consists in 5 lines of Python code).

\newpage
\subsection{A simple example}\label{case-study}

\begin{figure}[H]
  \begin{minipage}[c]{0.6\linewidth}
      \begin{lstlisting}[language=Python, linewidth=0.8\linewidth]
        from prose import Sequence, FitsManager
        from prose import blocks
          
        fm = FitsManager("images-folder")
        
        sequence = Sequence([
          blocks.Calibration(**fm.calibs),
          blocks.DAOFindStars(n=100),
          blocks.Moffat2D(),
          blocks.Get("fwhm"),
          blocks.Video("video.gif"),
          blocks.Stack(),
        ], fm.images)
        
        sequence.run()
      \end{lstlisting}
  \end{minipage}
  \begin{minipage}[c]{0.55\linewidth}
    \centering
	  \includegraphics[width=\linewidth]{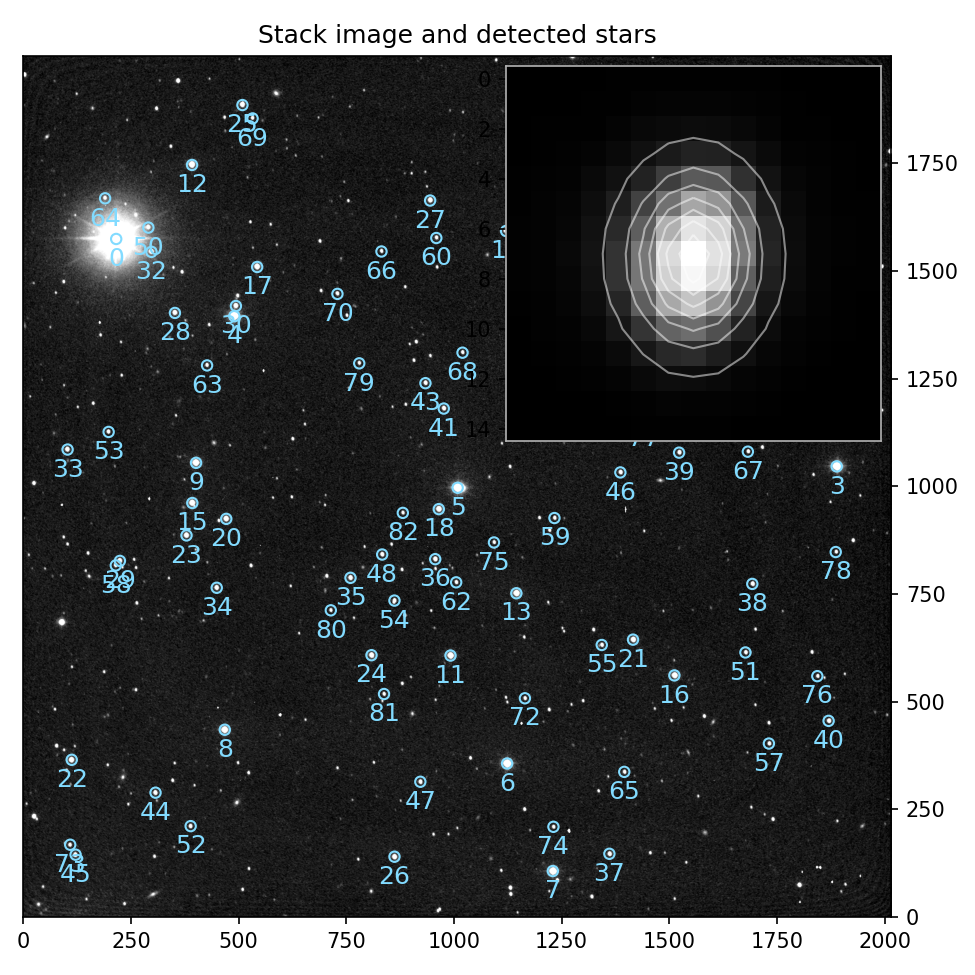}
  \end{minipage}
	\caption{(Left) Python code of the pipeline described in \autoref{case-study}, performing the characterization of images effective PSF over a full night of observation. (Right) resulting stack image and PSF model of the observation produced using products directly accessible from the \texttt{sequence} object. This example is made from 742 10s exposure images taken by the TRAPPIST-South telescope on December 14, 2018.}
  \label{img:case_study}
\end{figure}

Most of \prose{} \Blocks{} use state-of-the-art methods and libraries. Its novelty stands in its object-oriented structure, allowing for powerful abstraction in implementing custom pipelines. In this section, we present a simple example: a pipeline to characterize the effective point-spread-function (PSF) of images taken during a full night of observation. For every image, the pipeline starts with bias, dark and flat calibration \citep[section~4.5]{howell}. Then, bright stars are detected to build an effective PSF cutout and characterized using a specific PSF model. Here, we will use a 2D Moffat profile \citep{Moffat1969} and only focus on its full width at half maximum (FWHM). Finally, images are stacked and a movie of the night is generated for further vetting.
\bigskip\\
The Python code to build such a pipeline with \prose{} is shown in \autoref{img:case_study}, where \Blocks{} have been assembled into a single \Sequence{} to perform each task.
\bigskip\\

In this pipeline, a wide range of stellar profiles or detection techniques may be experimented with through the use of different \Blocks{}. The \prose{} online documentation\footnote{\prosegithub} provides many more examples as well as the description of several convenient tools, such as the \texttt{FitsManager} object, allowing to deal with highly disorganized FITS folders by using their header information.

\newpage
\section{Base photometric pipeline} \label{pipeline}

\subsection{Principle}
Using the structure described in \autoref{prose}, \prose{} features a standard photometric reduction pipeline which requires minimal user input and can be decomposed in four \Sequences{}: Sequence 1 consists in detecting the $n$ brightest stars from a single image taken from the full observation. These $n$ positions are stored and will act as a reference to align all other frames on. During Sequence 2, each image goes through bias, dark, and flat calibration \cite[section~4.5]{howell}. On each of them, we detect the $n$ brightest stars and compare their positions to the set of reference ones. Image misalignment from the reference can reasonably be approximated by an affine transformation that we compute using the method described in \citet{Lang_2010} (see \autoref{twirl}). We then form a stack image out of transformed frames using bi-linear interpolation but we only save the untransformed calibrated frames as to make the photometry on non-interpolated images. In sequence 3, we detect stars in the stack image (500 by default) and characterize their profile to aid in scaling apertures radii in the next sequence. Finally, sequence 4 consists in extracting the detected stars' fluxes using circular aperture photometry \citep{daophot} on calibrated frames. Although they are not aligned, stars positions in these images can be computed from the ones detected in the stack image (aligned to the reference image) by applying the inverse transformation matrix previously computed and stored for each exposure. As the detector sampling might introduce an error on these local coordinates, a centroiding algorithm is finally used (see \autoref{ballet}) to properly center each aperture on every image from which the fluxes are finally extracted. \autoref{tab:pipe} shows the four sequences described above: 1. Reference stars detection to align other images on; 2. Calibration, alignment and stacking of every frame; 3. A global star detection on the stack image; and 4. Aperture photometry on each calibrated frame refined by a centroiding algorithm.
\bigskip\\

The inverse transformation we rely on in our pipeline is indirectly adopted by many reduction tools (such as \citealt{Collins2017}) where stars positions are set in a World Coordinate System (WCS, \citealt{wcs}). Indeed, some of them require all frames to be plate-solved which is generally done using \citet{Lang_2010} through the \href{http://astrometry.net/}{astrometry.net} service. While we leverage the same asterisms matching algorithm (see \autoref{twirl}), we do not relate our transformation to a WCS and allow raw images to be processed in a standalone and efficient way. We note that \AIJ{} do not require plate-solved images and rely on a centroiding algorithm to reposition apertures on drifting stars. However, this strategy is only valid for images whose translation is lower than the aperture's radius, which is what motivated the addition of a complete registration algorithm in our pipeline. We also note that the apertures used in the aperture photometry task are scaled based on an analytical fit to the effective PSF of the detector, hence assuming a known PSF model. In its current version, \prose{} features \Blocks{} to account for Gaussian and Moffat profiles (\cite{Moffat1969}, see \autoref{ballet}), which are only suited to focused PSF. For this reason, follow-up work will include the development of \Blocks{} to properly treat unfocused profiles (which are often ring-shaped and currently depart from default settings).\bigskip\\

The products we obtain by running the pipeline can be accessed within a Python instance and saved for later use in a \texttt{.phot} file including the stacked frame of the observation, as well as the positions and fluxes of detected stars at all times and apertures given with their uncertainties. A variety of time-series recorded by individual \Blocks{} are saved as well, such as the mean FWHM of the effective PSF, airmass, background level or frame misalignment, overall essential to model the extent of systematic signals in the measured fluxes.

\begin{table}[H]
	\centering
	\bgroup
	\def\arraystretch{1.1}
	\begin{tabular}{ll}
	  \multicolumn{2}{l}{1. Reference stars detection (on the reference image)}\\
	  \hline
	  \texttt{Calibration} & Bias, dark, and flat calibration\\
	  \texttt{Trim} & Border trimming\\
	  \texttt{SegmentedPeaks} & Detection of a reference set of stars\\
	  \multicolumn{2}{l}{}\\
	  \multicolumn{2}{l}{2. Calibration, alignment and stacking (on all raw images)}\\
	  \hline
	  \texttt{Calibration} & Bias, dark and flat calibration\\
	  \texttt{Trim} & Border trimming\\
	  \texttt{Flip} & Image flip according to the reference\\
	  \texttt{SegmentedPeaks} & Fast detection of stars positions\\
	  \texttt{Moffat2D} & Effective PSF characterization\\
	  \texttt{Twirl} & Computing affine transformation\\
	  \texttt{SaveReduced} & Saving calibrated image\\
	  \texttt{AffineTransform} & Applying affine transformation\\
	  \texttt{Stack} & Adding transformed image to stack\\
	  \texttt{Video} & Adding transformed image to video\\
	  \multicolumn{2}{l}{}\\
	  \multicolumn{2}{l}{3. Stars detection (on the stack image)}\\
	  \hline
	  \texttt{DAOFindStars} & Stars detection\\
	  \texttt{Moffat2D} & Stacked effective PSF characterization\\
	  \multicolumn{2}{l}{}\\
	  \multicolumn{2}{l}{4. Aperture photometry (on calibrated images)}\\
	  \hline
	  \texttt{AffineTransform} & Applying inverse affine transformation\\
	  \texttt{BalletCentroid} & Refining stars centroid positions\\
	  \texttt{PhotutilsAperturePhotometry} & Aperture photometry\\
	  \texttt{XArray} & Saving fluxes and other measurements. \\
  
	\end{tabular} 
  
	\egroup
  
	\vspace{0.5cm}
	\caption{The four sequences of \prose{} default photometric pipeline}
	\label{tab:pipe}
\end{table}

\subsection{Differential photometry}\label{diff-lc}

During the reduction, visible stars are automatically detected for every frame and aperture photometry is performed over a range of 40 apertures, yielding ideal measurements to leverage the power of ensemble differential photometry. For that purpose, \prose{} implements a pure Python version of the algorithm described in \citet{Broeg2005}. This algorithm consists in building an \textit{artificial comparison star} using the weighted sum of all available stars in the field. Appropriate weights are iteratively found as to minimize variance over all differential light curves. Doing so, higher weights are given to stars displaying lower variability and higher signal-to-noise ratio (SNR), more likely to feature systematic signals shared among all sources. Finally, the best aperture is chosen as the one minimizing the white noise of the target's light curve, estimated with the median standard deviation of points within 5 minutes bins. As all products are saved within a \texttt{.phot}  file, light curves for different apertures can later be explored and the best one selected manually.
\bigskip\\
Our photometric pipeline relies on a strategy which consists in registering all images before refining stars centroids to perform aperture photometry. If processing time is a concern, general registration and centroiding algorithms can be a problem. To keep it fast on domestic computers, we rely on two methods, integrated into \Blocks{}, yielding fast and robust estimates: \texttt{Twirl} which computes the affine transformation between two sets of stars using the method from \citet{Lang_2010}, and \texttt{BalletCentroid}, a novel centroiding algorithm based on the convolutional neural network presented in \citet{Herbel2018}. We present these methods in the next sections.

\subsection{\texttt{Twirl}: stars registration}\label{twirl}
\begin{figure}
	\centering
	\includegraphics[width=0.8\linewidth]{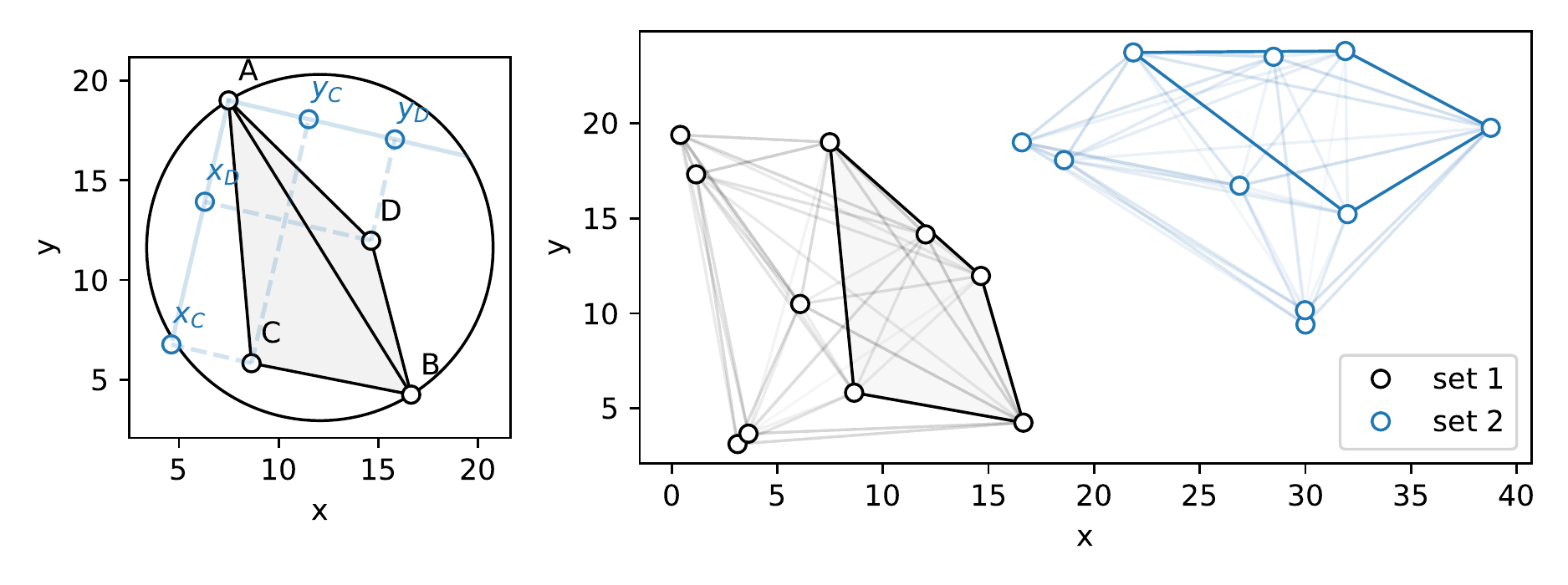}
	\caption{Demonstration of the four-points asterisms used in Twirl. The left figure is reproduced from \citet{Lang_2010} and shows how a unique hash code $(x_C, y_C, x_D, y_D)$ is formed from four points coordinates: by taking the relative positions of two of the coordinates (C, D) with respect to the outermost ones (A and B). The right plot shows the matching of this same asterism with one from another set.}
	\label{fig:twirl}
  \end{figure}

Stars registration is a common need in astronomy. Given two sets of stars positions, it consists in identifying pairs of coordinates corresponding to unique objects, but observed from different referential, hence giving an absolute reference in the sky. To do that, many algorithms, as well as humans, birds, and seals \citep{Foster2018}, use \textit{asterisms}: shapes produced by stars arrangements, like constellations, recognizable in the sky. In our case, we will use the method of  \citet{Lang_2010} standing at the core of the widely adopted astrometric calibration service \href{http://astrometry.net/}{astrometry.net}. While it produces robust astrometric solutions, it requires a cross-match to reference stars catalogs that we want to avoid in our pipeline. Hence, our implementation does not provide a complete WCS solution but a transformation matrix allowing stars to be registered to a common frame.
\bigskip\\
The algorithm starts by computing four-points asterisms, translated into unique hash codes consisting in four numbers  as described in \autoref{fig:twirl}. These codes have the advantage of being invariant over any affine transformation. Then, starting with two sets of coordinates, \textit{set 1} and \textit{set 2}, their hash codes can be produced and matched to find the transformation they originate from. We produce unique hash codes the same way as \citet{Lang_2010} but we use a simpler matching strategy. We organize the hash codes into a kd-tree structure (as in \citet{Lang_2010} but using an implementation from \texttt{scipy.spatial.KDTree}\footnote{\href{https://docs.scipy.org/doc/scipy/reference/generated/scipy.spatial.KDTree.html}{https://docs.scipy.org/doc/scipy/reference/generated/scipy.spatial.KDTree.html}}) which allows for fast nearest-neighbor lookup. For all asterims in \textit{set 1}, we retrieve the closest asterism from \textit{set 2}, compute its corresponding transformation, apply it to \textit{set 1} and count the number of stars that can be cross-matched between the two sets (given a tolerance). Finally, we keep the transformation maximizing the number of matched stars.\\
  
Our method is implemented in the \texttt{Twirl} \Block{} which, given reference coordinates, computes and stores the affine transformation matrix from one set to the reference one. We make its Python implementation (separated from \prose{}) publicly available in the \texttt{twirl} repository \footnote{\href{https://github.com/lgrcia/twirl}{https://github.com/lgrcia/twirl}}.

\subsection{\texttt{ballet}: Fast centroiding}\label{ballet}

Stars positions being computed, as well as their alignment to a common reference, the caveat of our strategy is to rely on a centroiding algorithm to refine their centroids on every frame, which can be CPU intensive and lead to variable results depending on the method used. Inspired by \cite{Herbel2018}, we trained a deep Convolutional Neural Network (CNN) to predict accurate centroid positions out of normalized stars cutouts.
\\

\begin{figure}
  \centering
  \includegraphics[width=0.7\linewidth]{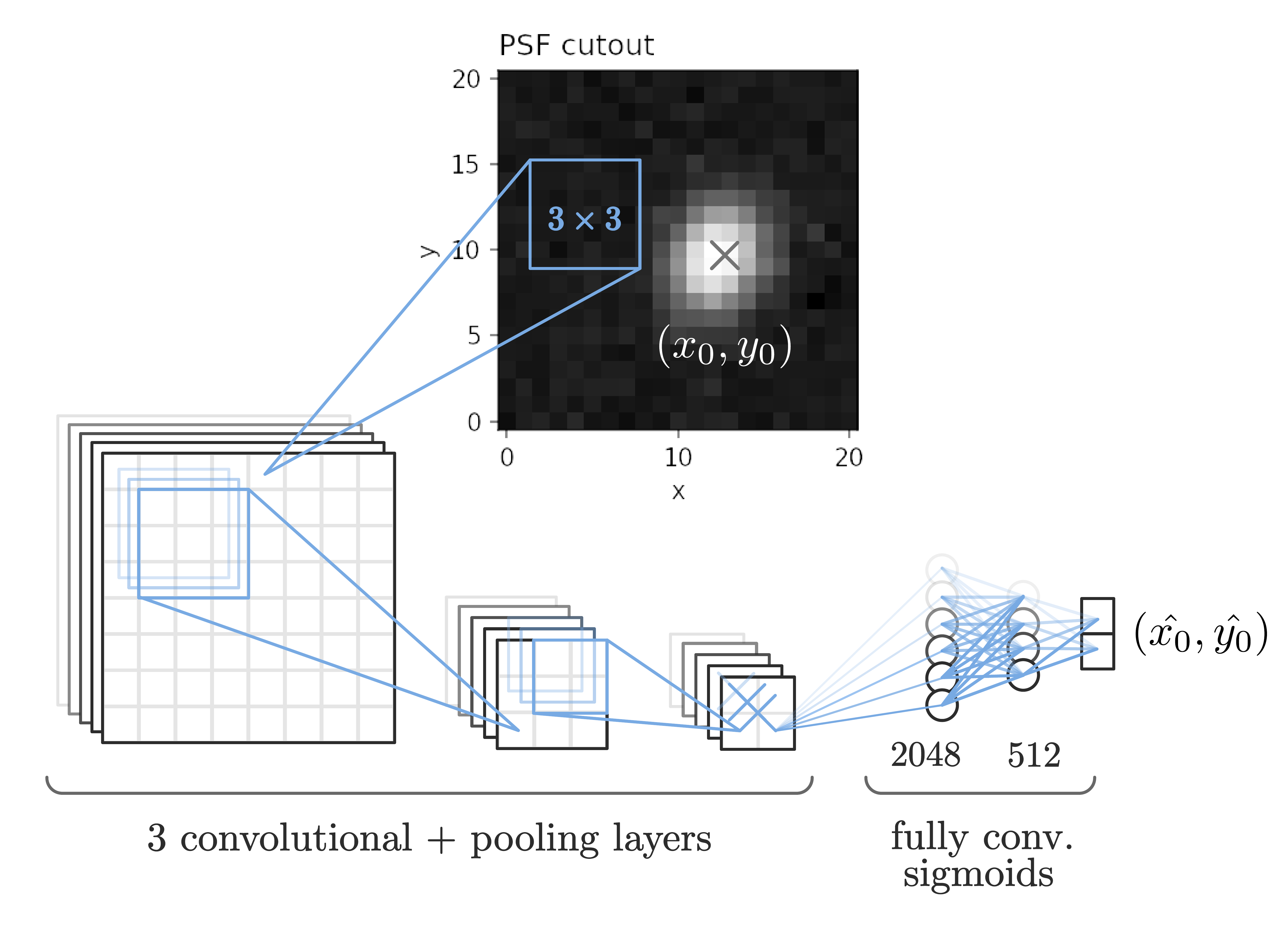}
  \caption{Neural network architecture}
  \label{fig:nn_architecture}
\end{figure}


The architecture of our neural network is shown in \autoref{fig:nn_architecture} and consists in 3 ReLu-activated convolutional layers of sizes 64, 128 and 256, each followed by a pooling layer downsampling their outputs by a factor 2. The 1028 resulting weights serve as inputs to 2 sigmoid-activated fully connected layers, finally leading to the centroid $(x, y)$ coordinates. This CNN was implemented with the \texttt{tensorflow} package, and trained using the Adam stochastic gradient descent algorithm (\citealt{kingma2017adam}) on a Huber loss function \citep{huber}. The training samples consisted in $2\cdot10^5$ simulated Moffat profiles \cite{Moffat1969} expressed as:

\begin{gather}
  F(x, y) = b + \frac{1}{\left(1 + \frac{dx^2}{\sigma_x^2} + \frac{dy^2}{\sigma_y^2}\right)^\beta} \\ \\
  \text{where} \begin{cases}
      dx = (x-x_0)\cos{\theta} + (y-y_0)\sin{\theta}\\
      dy = -(x-x_0)\sin{\theta} + (y-y_0)\cos{\theta}\\
    \end{cases} 
\end{gather}
\\

with $(x_0, y_0)$ the centroid position, $(\sigma_x, \sigma_y)$ the standard deviation of the point-spread-function (PSF) in the $x$ and $y$-axis, $\theta$ its relative angle with respect to the $x$-axis, $\beta$ the concentration of the Moffat profile and $b$ its relative background. To train the model, these parameters were randomly drawn from the distributions given in \autoref{tab:nn_dist}, representative of a wide range of PSF characteristics, including widely off-center ones. We note that this training dataset is only representative of focused PSF. 
\bigskip\\ 

\begin{table}[H]
  \centering
  \bgroup
  \def\arraystretch{1.3}
  \begin{tabular}{cc}
	\hline
    Parameter & Distribution\\
    \hline
    $x_0, y_0$ & $\mathcal{U}(3, 12)$\\
    $\sigma_x$  & $\mathcal{U}(2.5, 7.5)$\\
    $\sigma_y$  & $\sigma_x \cdot \mathcal{U}(0.5, 1.5)$\\
    $\theta$  & $\mathcal{U}(0, \pi/8) $\\
    $\beta$ & $\mathcal{U}(1, 8)$\\
    $b$ & $\mathcal{U}(0, 0.1)$\\
	\hline
  \end{tabular} 
  \egroup
  \vspace{0.5cm}
  \caption{Training sample parameters distributions. $\mathcal{U}(a,b)$ is a uniform distribution bounded by $a$ and $b$.}
  \label{tab:nn_dist}
\end{table}

To assess the performances of our trained network, we draw another $10^4$ samples from the same distributions and compare our predictions to three other centroiding methods. The first consists in fitting the star cutout with a two-dimensional quadratic polynomial; the second is a fit of the marginal $x$ and $y$ PSF to univariate Gaussians; and the third is a complete fit of the cutout to a two-dimensional Gaussian. These are implemented in the \texttt{photutils}\footlink{https://photutils.readthedocs.io/en/stable/centroids.html} package \citep{larry_bradley_2020_4044744} under the name \texttt{centroid\_quadratic}, \texttt{centroid\_1dg} and \texttt{centroid\_2dg}, that we employ to make our comparison. We show that our optimized neural network, named \texttt{ballet}, predicts centroid positions with an error of 0.015 pixels, comparable to the precision obtained from directly fitting a two-dimensional Gaussian while being two orders of magnitudes faster (see \autoref{fig:nnperf}). Again, we highlight that our model is suited and trained only for focused PSF, and plan to extend it to defocused stellar profiles in a follow-up work.

\begin{figure}[H]
\centering
  \bgroup
    \def\arraystretch{1.1}
    \begin{tabular}[b]{|l|c|c|c|}
      \hline
      \multirow{2}{*}{method} & \multirow{2}{*}{\shortstack[c]{accuracy \\ ($10^{-2}\cdot pix.)$}} & \multirow{2}{*}{\shortstack[c]{time ($s$)}} & \multirow{2}{*}{robustness} \\
      & & & \\
      \hline
      Quadratic & 4.132 & 0.41 & 92.8 \%\\
      1D Gaussian & 2.271 & 39.50 & 99.5 \%\\
      2D Gaussian & 1.516 & 70.23 & 99.9 \% \\
      \texttt{ballet} & 1.524 & 0.33 & 99.8 \%\\
      \hline
    \end{tabular} 
    \egroup
  \caption{Performance assessment of multiple centroiding algorithms against our method using $10^4$ samples drawn from the distributions described in  \autoref{tab:nn_dist}. In the right table: \textit{accuracy} corresponds to the root-mean-square of the differences between the true centroid coordinates and their predicted values; \textit{time} is the processing time to predict the positions of 1000 PSF cutouts; and \textit{robustness} is the percentage of errors under 0.1 pixels, which is linked to the occurrence of prediction outliers. Our method is referred to as \texttt{ballet}.}
  \label{fig:nnperf}
\end{figure}

For more details about this particular application of CNNs, we refer the interested readers to \cite{Herbel2018} in which we found rich references. Choosing and tweaking the architecture of a neural network yet remains empirical and ours is the result of many manual optimizations. We make this pre-trained model and its \texttt{tensorflow}\footnote{\href{https://www.tensorflow.org/}{https://www.tensorflow.org/}} implementation available in the \texttt{ballet} repository\footnote{\href{https://github.com/lgrcia/ballet}{https://github.com/lgrcia/ballet}}, which we intend to complement with other CNN applications in the future. Most importantly we integrate the model presented in this section into the \texttt{BalletCentroid} \textit{Block}, which makes its use straightforward in \prose{} photometric pipeline and any other developed under the \prose{} architecture.

\newpage
\section{Comparison with Astroimagej: Tess follow-ups} \label{aij}

The power of the pipeline described in \autoref{pipeline} is to rely on general-purpose \Blocks{} leading to rich sets of measurements to work on. In this section, we present a comparison of this pipeline with \AstroImageJ{} (\AIJ{}, \citealt{Collins2017}), a graphical user interface to process astronomical images into high-precision photometry. \AIJ{} has become a reference in the field of transiting exoplanet observations and a standard for the TFOP activities. For this reason, we will do our comparison on \aijobs TESS follow-up observations conducted from the TRAPPIST-South telescope between July 2018 and March 2021 (see \autoref{tab:aij-dataset}). We note that this dataset contains only focused observations, well suited for the default settings of \prose{} photometric reduction pipeline. 

\subsection{\prose{} reductions}
\prose{} reductions were done in bulk and using the default settings of its photometric reduction pipeline, the only interaction from the user being the manual selection of the target star on the stack images. As required by our alignment method, the default number of reference stars $n$ is set to 12 (generally providing a good balance between the accuracy and speed of the \texttt{Twirl} method described in \autoref{twirl}) and the reference image is chosen as the one from the middle of the observation. Finally, the differential light curves were automatically built using \cite{Broeg2005} as described in \autoref{diff-lc}.

\subsection{\AstroImageJ{} reductions}
The photometric reduction done with \AIJ{} follows the TFOP SG1 guidelines and the good practices described in Conti, D. M. (2018)\footlink{https://astrodennis.com/Guide.pdf}. First, the image reduction was done using the \textit{Data Reduction Facility}, following calibration steps similar to the ones described in \autoref{pipeline}. Once all images calibrated and aligned, the target star was identified and the seeing profile plotted. The \textit{Measurement tool} was used to select comparison stars of brightnesses comparable to that of the selected target, with a magnitude difference below 1 for crowded fields, and below 1.5 for scarcer fields. The aperture photometry was then performed on the calibrated images using an aperture adjusted to the seeing profile. The final set of comparison stars was manually selected as the one yielding the lower white noise for the target's light curve. Finally, this operation was repeated and optimized at different apertures, the final one selected as the one yielding the lower white and red noise for the target's light curve. As our objective is to compare the differential fluxes obtained from both reductions, we did not rely on the rich set of modeling tools offered by \AIJ{}.

\subsection{Metrics}\label{metrics}
The performance of both tools is assessed using four distinct metrics. The first is the median standard deviation of the target light curve within 12 points bins (arbitrary), a proxy to the white noise variance. The second and third metric will be the white and red noise computed using a method inspired from \cite{Pont2006} which has the following principle: if the light curve only contains white noise, the variance of the binned light curve with bins of varying sizes would be inversely proportional to the number of points in the bins. In opposition, the variance of varying-sized bins will be independent of the bin size for a light curve with a purely correlated signal. The white and red noise can then be obtained by fitting the bin-size dependent variance of the light curve to:

\begin{equation}
	\nu(n) = \frac{\sigma_w^2}{n} + \sigma_r^2
\end{equation}

where $\sigma_w$ and $\sigma_r$ are the white and red noise standard deviations and $n$ the bins size. The last metric we consider is the single-transit SNR as described in \cite{Pont2006}, i.e. : 

\begin{equation}\label{eq:snr}
  S_r= \frac{df}{\sqrt{\frac{\sigma_w^2}{n} + \sigma_r^2}}
\end{equation}

where $df$ is the relative depth of the transit and $n$ is the number of points in transit. Most of the light curves of our sample contain a transit signal and a meridian flip\footnote{A flip of the telescope required to track objects passing across the meridian line in the case of German equatorial mounts.} that, if being ignored, could be considered as red noise. To avoid that, the estimation of $\sigma_r$ is done on light curves for which the meridian flip has been modeled and removed, simultaneously with an empirical transit model from \cite{protopapas} whose maximum a posteriori parameters are found using the BFGS optimization algorithm \citep{bfgs}. We note that all light curves are correlated with airmass and that this correlation is strongly driven by the choice and weights of comparison stars. Since prose and \AIJ{} light curves are built on different sets of comparison stars, we remove these signals from all light curves by fitting an order two polynomial of time, simultaneously to the meridian flip and transit signal. Finally, we visually inspect all light curves (\autoref{fig:all_lcs}) and evaluate the four metrics presented in this section on differential fluxes cleaned from transit, meridian flip, and airmass signals.

\subsection{Results}

\begin{figure}
  \centering
  \includegraphics[width=0.7\linewidth]{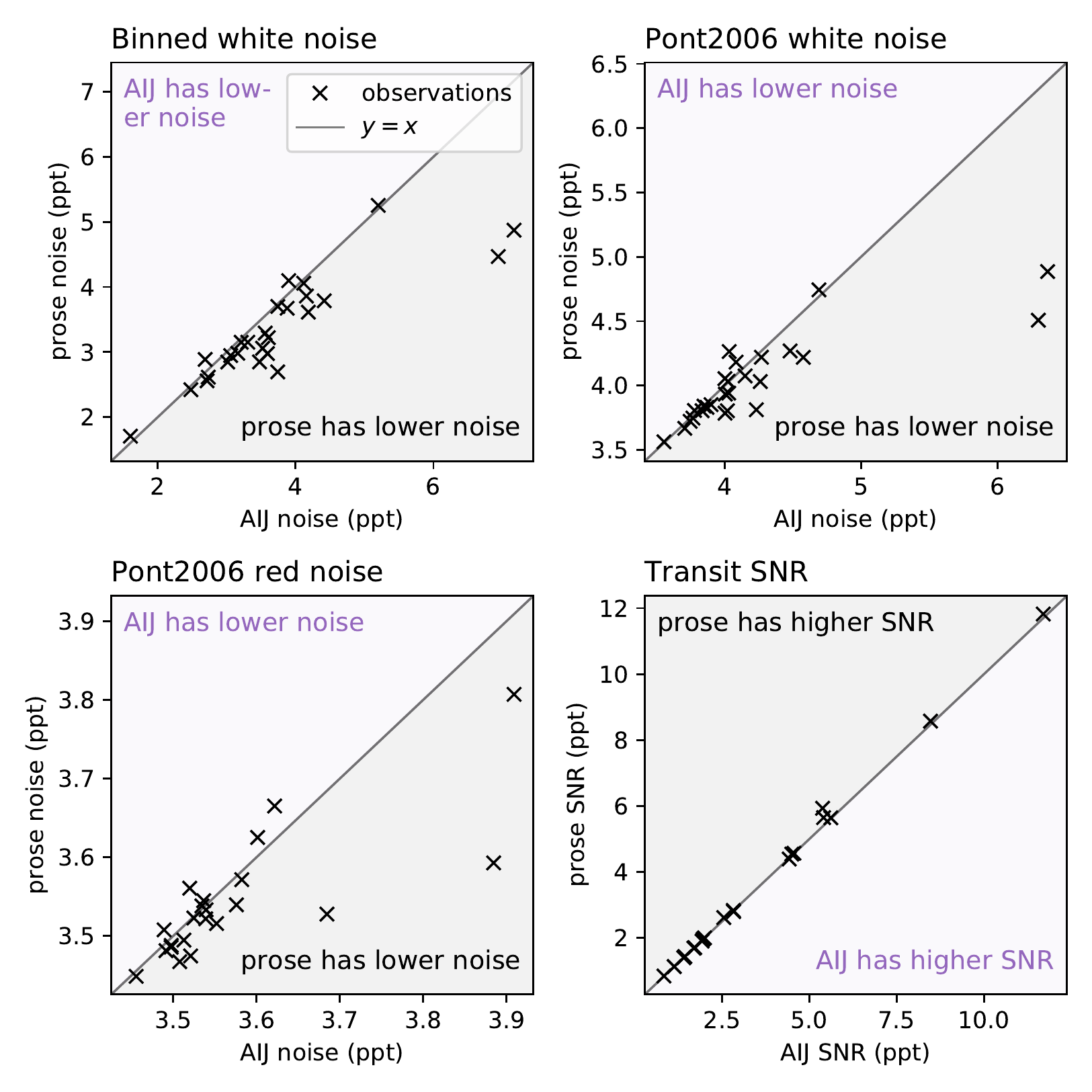}
  \caption{Noise and transit SNR comparison of \prose{} against \AstroImageJ{} (\AIJ{}) on \aijobs TESS follow-up observations.}
  \label{fig:aijcompare}
\end{figure}

\autoref{fig:aijcompare} shows the results of our comparison: \prose{} automatically yields light curves with lower white and red noise, resulting in transits recovered at slightly higher SNR on average. In this figure, the most discrepant points correspond to TOI-674, TOI-742, and TOI-2322 light curves (indexed 17, 24 and 26 in \autoref{tab:aij-dataset} and \autoref{fig:all_lcs}). A careful reanalysis of these three datasets with \AIJ{} was performed but could not explain these discrepancies.
\\

\AIJ{} does not require prior programming knowledge, making it a popular choice for professional astronomers as well as in the amateur astronomy community. Then, doing aperture photometry with \AIJ{} requires the user to manually pick comparison stars (up to version 4)  as well as a unique aperture size. In opposition, using standard \Blocks{} in \prose{} photometric pipeline allows to automatically detect stars, use a wide range of apertures and apply an automatic differential photometry algorithm, overall yielding light curves with lower noise. We note that, apart from alignment and centroiding aspects, \AIJ{} and \prose{} use similar algorithms (such as aperture photometry) which naturally leads to very comparable results. This validates the performances of \prose{} default photometric pipeline against \AIJ{}. 

\subsection{Processing time}
The \AIJ{} products used in our comparison were provided by several users with no information on the time required to do their complete reduction. In addition, \prose{} provides measurements on a much larger number of stars and a wide range of apertures, making the comparison with \AIJ{} uneven. That said, \prose{} pipeline was developed with domestic computers in mind, an aspect which deserves to be quantified.
\bigskip\\
Given an observation, the reduction time of \prose{} photometric pipeline depends on the number of images to process, their sizes, the number of stars to extract photometry on, the number of apertures considered, their radii, and of course the computer being used. Providing a figure for all these parameters is beyond the scope of this paper, instead we provide a figure which can be easily adapted to estimate the time required to reduce a given observation. Our sample contains $2096 \times 2080 $ pixels images on which we consider the 100 brightest stars using 40 apertures. We performed our reductions on a 3 Ghz Intel core i5 processor and 8Gb of RAM, taking an average of 10 $ms$ per image per star, which translates into 8 minutes for a typical 500 frames observation with photometry extracted on 100 stars and 40 apertures.

\subsection{Reports}
TESS candidates follow-ups are meant to be reported to the TFOP community. We wished to provide advanced reporting capabilities and developed a reporting framework adapted to \prose{} products. Our reports consist in \LaTeX{} templates that can be instantiated and automatically filled from Python scripts. While reports contain general sections, some TESS-specific pages were added, providing all the information related to TFOP activities. An example of such report is provided in \autoref{apreports} but its complete description is beyond the scope of this paper. For that, we refer the interested readers to the \prose{} online documentation\footnote{\prosegithub}.
\\

\newpage
\section{Conclusion}
Initiatives like  Astropy \citep{astropy:2013,astropy:2018} and its affiliated packages (such as \texttt{photutils} \citep{larry_bradley_2020_4044744}) succeeded in offering an alternative to community tools like IRAF \citep{iraf}, whose support and use is progressively fading. However, building and testing a complete image processing pipeline using these tools requires software development skills and time, and often leads to poorly maintainable solutions.
\bigskip\\
We presented \prose, a Python package to build modular image processing pipelines out of reusable \Blocks{}. To demonstrate its potential, we used it to develop a basic photometric reduction pipeline that we compared to the one of \AstroImageJ{} \citep{Collins2017}, a reference software at the center of the TESS candidates follow-up effort. Hence, we focused our comparison on \aijobs TESS follow-up observations and showed that \prose, modular and instrument-agnostic, led to light curves with reduced white and red noise compared to the ones produced with \AstroImageJ{}. Beyond its gain in precision, our pipeline requires close to no user interaction (apart from target selection) and leads to products benefiting from the advanced reporting capabilities of \prose, all directly exploitable in Python.
\bigskip\\
Applications of \prose{} extend beyond its base photometric pipeline. As an example, it was recently used to reduce slitless spectroscopy data from Hubble Space Telescope WFC3 observations (Garcia et al., 2021 in prep.). This development led to \Blocks{}, independently maintainable, that will soon be available to the community, potentially useful in other applications. By adopting this modular approach, we hope to see the development of transparent pipelines which can live beyond individuals projects, serving the entire community, and leading to reproducible results. 
\bigskip\\
We make \prose{} open-source under the MIT license in a version controlled repository\footlink{https://github.com/lgrcia/prose}.

\section*{Acknowledgements}
TRAPPIST is funded by the Belgian Fund for Scientific Research (Fond National de la Recherche Scientifique, FNRS) under the grant FRFC 2.5.594.09.F, with the participation of the Swiss National Science Foundation (SNF). MG and EJ are F.R.S.-FNRS Senior Research Associates. LD is F.R.S.-FNRS Postdoctoral Researcher. This publication benefits from the support of the French Community of Belgium in the context of the FRIA Doctoral Grant awarded to M.T. This work has received funding from the Balzan Prize Foundation and from the University of Liège. 

\section*{Data availability}
The example data used in \autoref{case-study} are privately held by the TRAPPIST team and can be requested by contacting the authors. On the other hand, some of the TESS follow-up observations used in this paper are available through the ExoFOP-TESS webpage\footlink{https://exofop.ipac.caltech.edu/tess/}. Raw images from the TRAPPIST telescope (listed in \autoref{tab:aij-dataset}), as well as their photometric products (\prose{} and/or \AIJ{}) can be requested to the authors.

\bibliography{bib} 

\begin{thebibliography}{}
\expandafter\ifx\csname natexlab\endcsname\relax\def\natexlab#1{#1}\fi
\providecommand{\url}[1]{\href{#1}{#1}}
\providecommand{\dodoi}[1]{doi:~\href{http://doi.org/#1}{\nolinkurl{#1}}}
\providecommand{\doeprint}[1]{\href{http://ascl.net/#1}{\nolinkurl{http://ascl.net/#1}}}
\providecommand{\doarXiv}[1]{\href{https://arxiv.org/abs/#1}{\nolinkurl{https://arxiv.org/abs/#1}}}

\bibitem[{{Astropy Collaboration} {et~al.}(2013){Astropy Collaboration},
  {Robitaille}, {Tollerud}, {Greenfield}, {Droettboom}, {Bray}, {Aldcroft},
  {Davis}, {Ginsburg}, {Price-Whelan}, {Kerzendorf}, {Conley}, {Crighton},
  {Barbary}, {Muna}, {Ferguson}, {Grollier}, {Parikh}, {Nair}, {Unther},
  {Deil}, {Woillez}, {Conseil}, {Kramer}, {Turner}, {Singer}, {Fox}, {Weaver},
  {Zabalza}, {Edwards}, {Azalee Bostroem}, {Burke}, {Casey}, {Crawford},
  {Dencheva}, {Ely}, {Jenness}, {Labrie}, {Lim}, {Pierfederici}, {Pontzen},
  {Ptak}, {Refsdal}, {Servillat}, \& {Streicher}}]{astropy:2013}
{Astropy Collaboration}, {Robitaille}, T.~P., {Tollerud}, E.~J., {et~al.} 2013,
  \aap, 558, A33, \dodoi{10.1051/0004-6361/201322068}

\bibitem[{{Astropy Collaboration} {et~al.}(2018){Astropy Collaboration},
  {Price-Whelan}, {Sip{\H{o}}cz}, {G{\"u}nther}, {Lim}, {Crawford}, {Conseil},
  {Shupe}, {Craig}, {Dencheva}, {Ginsburg}, {Vand erPlas}, {Bradley},
  {P{\'e}rez-Su{\'a}rez}, {de Val-Borro}, {Aldcroft}, {Cruz}, {Robitaille},
  {Tollerud}, {Ardelean}, {Babej}, {Bach}, {Bachetti}, {Bakanov}, {Bamford},
  {Barentsen}, {Barmby}, {Baumbach}, {Berry}, {Biscani}, {Boquien}, {Bostroem},
  {Bouma}, {Brammer}, {Bray}, {Breytenbach}, {Buddelmeijer}, {Burke},
  {Calderone}, {Cano Rodr{\'\i}guez}, {Cara}, {Cardoso}, {Cheedella}, {Copin},
  {Corrales}, {Crichton}, {D'Avella}, {Deil}, {Depagne}, {Dietrich}, {Donath},
  {Droettboom}, {Earl}, {Erben}, {Fabbro}, {Ferreira}, {Finethy}, {Fox},
  {Garrison}, {Gibbons}, {Goldstein}, {Gommers}, {Greco}, {Greenfield},
  {Groener}, {Grollier}, {Hagen}, {Hirst}, {Homeier}, {Horton}, {Hosseinzadeh},
  {Hu}, {Hunkeler}, {Ivezi{\'c}}, {Jain}, {Jenness}, {Kanarek}, {Kendrew},
  {Kern}, {Kerzendorf}, {Khvalko}, {King}, {Kirkby}, {Kulkarni}, {Kumar},
  {Lee}, {Lenz}, {Littlefair}, {Ma}, {Macleod}, {Mastropietro}, {McCully},
  {Montagnac}, {Morris}, {Mueller}, {Mumford}, {Muna}, {Murphy}, {Nelson},
  {Nguyen}, {Ninan}, {N{\"o}the}, {Ogaz}, {Oh}, {Parejko}, {Parley}, {Pascual},
  {Patil}, {Patil}, {Plunkett}, {Prochaska}, {Rastogi}, {Reddy Janga},
  {Sabater}, {Sakurikar}, {Seifert}, {Sherbert}, {Sherwood-Taylor}, {Shih},
  {Sick}, {Silbiger}, {Singanamalla}, {Singer}, {Sladen}, {Sooley},
  {Sornarajah}, {Streicher}, {Teuben}, {Thomas}, {Tremblay}, {Turner},
  {Terr{\'o}n}, {van Kerkwijk}, {de la Vega}, {Watkins}, {Weaver}, {Whitmore},
  {Woillez}, {Zabalza}, \& {Astropy Contributors}}]{astropy:2018}
{Astropy Collaboration}, {Price-Whelan}, A.~M., {Sip{\H{o}}cz}, B.~M., {et~al.}
  2018, \aj, 156, 123, \dodoi{10.3847/1538-3881/aabc4f}

\bibitem[{{Bertin} \& {Arnouts}(1996)}]{sextractor}
{Bertin}, E., \& {Arnouts}, S. 1996, \aaps, 117, 393,
  \dodoi{10.1051/aas:1996164}

\bibitem[{Bradley {et~al.}(2020)Bradley, Sip{\H o}cz, Robitaille, Tollerud,
  Vin{\'{\i}}cius, Deil, Barbary, Wilson, Busko, G{\"u}nther, Cara, Conseil,
  Bostroem, Droettboom, Bray, Bratholm, Lim, Barentsen, Craig, Pascual, Perren,
  Greco, Donath, de~Val-Borro, Kerzendorf, Bach, Weaver, D'Eugenio, Souchereau,
  \& Ferreira}]{larry_bradley_2020_4044744}
Bradley, L., Sip{\H o}cz, B., Robitaille, T., {et~al.} 2020, astropy/photutils:
  1.0.0, 1.0.0,  Zenodo, \dodoi{10.5281/zenodo.4044744}

\bibitem[{{Broeg} {et~al.}(2005){Broeg}, {Fern{\'a}ndez}, \&
  {Neuh{\"a}user}}]{Broeg2005}
{Broeg}, C., {Fern{\'a}ndez}, M., \& {Neuh{\"a}user}, R. 2005, Astronomische
  Nachrichten, 326, 134, \dodoi{10.1002/asna.200410350}

\bibitem[{{Collins} {et~al.}(2017){Collins}, {Kielkopf}, {Stassun}, \&
  {Hessman}}]{Collins2017}
{Collins}, K.~A., {Kielkopf}, J.~F., {Stassun}, K.~G., \& {Hessman}, F.~V.
  2017, \aj, 153, 77, \dodoi{10.3847/1538-3881/153/2/77}

\bibitem[{{Foreman-Mackey}(2016)}]{corner}
{Foreman-Mackey}, D. 2016, The Journal of Open Source Software, 1, 24,
  \dodoi{10.21105/joss.00024}

\bibitem[{{Foreman-Mackey} {et~al.}(2021){Foreman-Mackey}, {Luger}, {Agol},
  {Barclay}, {Bouma}, {Brandt}, {Czekala}, {David}, {Dong}, {Gilbert},
  {Gordon}, {Hedges}, {Hey}, {Morris}, {Price-Whelan}, \& {Savel}}]{exoplanet}
{Foreman-Mackey}, D., {Luger}, R., {Agol}, E., {et~al.} 2021, arXiv e-prints,
  arXiv:2105.01994.
\newblock \doarXiv{2105.01994}

\bibitem[{Foster {et~al.}(2018)Foster, Smolka, Nilsson, \& Dacke}]{Foster2018}
Foster, J.~J., Smolka, J., Nilsson, D.-E., \& Dacke, M. 2018, Proceedings of
  the Royal Society B: Biological Sciences, 285, 20172322,
  \dodoi{10.1098/rspb.2017.2322}

\bibitem[{{Gillon} {et~al.}(2011){Gillon}, {Jehin}, {Magain}, {Chantry},
  {Hutsem{\'e}kers}, {Manfroid}, {Queloz}, \& {Udry}}]{trapmg}
{Gillon}, M., {Jehin}, E., {Magain}, P., {et~al.} 2011, in European Physical
  Journal Web of Conferences, Vol.~11, European Physical Journal Web of
  Conferences, 06002, \dodoi{10.1051/epjconf/20101106002}

\bibitem[{Head \& Zerner(1985)}]{bfgs}
Head, J.~D., \& Zerner, M.~C. 1985, Chemical Physics Letters, 122, 264,
  \dodoi{https://doi.org/10.1016/0009-2614(85)80574-1}

\bibitem[{{Herbel} {et~al.}(2018){Herbel}, {Kacprzak}, {Amara}, {Refregier}, \&
  {Lucchi}}]{Herbel2018}
{Herbel}, J., {Kacprzak}, T., {Amara}, A., {Refregier}, A., \& {Lucchi}, A.
  2018, \jcap, 2018, 054, \dodoi{10.1088/1475-7516/2018/07/054}

\bibitem[{{Howell}(2006)}]{howell}
{Howell}, S.~B. 2006, {1}, Vol.~5, {Handbook of CCD Astronomy}

\bibitem[{Huber(1964)}]{huber}
Huber, P.~J. 1964, The Annals of Mathematical Statistics, 35, 73 ,
  \dodoi{10.1214/aoms/1177703732}

\bibitem[{{Jehin} {et~al.}(2011){Jehin}, {Gillon}, {Queloz}, {Magain},
  {Manfroid}, {Chantry}, {Lendl}, {Hutsem{\'e}kers}, \& {Udry}}]{Trappist}
{Jehin}, E., {Gillon}, M., {Queloz}, D., {et~al.} 2011, The Messenger, 145, 2

\bibitem[{Kingma \& Ba(2017)}]{kingma2017adam}
Kingma, D.~P., \& Ba, J. 2017, Adam: A Method for Stochastic Optimization.
\newblock \doarXiv{1412.6980}

\bibitem[{Lang {et~al.}(2010)Lang, Hogg, Mierle, Blanton, \&
  Roweis}]{Lang_2010}
Lang, D., Hogg, D.~W., Mierle, K., Blanton, M., \& Roweis, S. 2010, The
  Astronomical Journal, 139, 1782–1800, \dodoi{10.1088/0004-6256/139/5/1782}

\bibitem[{{Moffat}(1969)}]{Moffat1969}
{Moffat}, A.~F.~J. 1969, \aap, 3, 455

\bibitem[{{Pont} {et~al.}(2006){Pont}, {Zucker}, \& {Queloz}}]{Pont2006}
{Pont}, F., {Zucker}, S., \& {Queloz}, D. 2006, \mnras, 373, 231,
  \dodoi{10.1111/j.1365-2966.2006.11012.x}

\bibitem[{{Protopapas} {et~al.}(2005){Protopapas}, {Jimenez}, \&
  {Alcock}}]{protopapas}
{Protopapas}, P., {Jimenez}, R., \& {Alcock}, C. 2005, \mnras, 362, 460,
  \dodoi{10.1111/j.1365-2966.2005.09305.x}

\bibitem[{{Stetson}(1987)}]{daophot}
{Stetson}, P.~B. 1987, \pasp, 99, 191, \dodoi{10.1086/131977}

\bibitem[{{Tody}(1986)}]{iraf}
{Tody}, D. 1986, in Society of Photo-Optical Instrumentation Engineers (SPIE)
  Conference Series, Vol. 627, Instrumentation in astronomy VI, ed. D.~L.
  {Crawford}, 733, \dodoi{10.1117/12.968154}

\bibitem[{{Wells} \& {Greisen}(1979)}]{wcs}
{Wells}, D.~C., \& {Greisen}, E.~W. 1979, in Image Processing in Astronomy, ed.
  G.~{Sedmak}, M.~{Capaccioli}, \& R.~J. {Allen}, 445

\end{thebibliography}

\appendix
\section{Observations}

\subsection{\AIJ{} comparison dataset}\label{aij-dataset}
\begin{figure}[H]
  \hspace{-1.4cm}
  \includegraphics[width=1.1\linewidth]{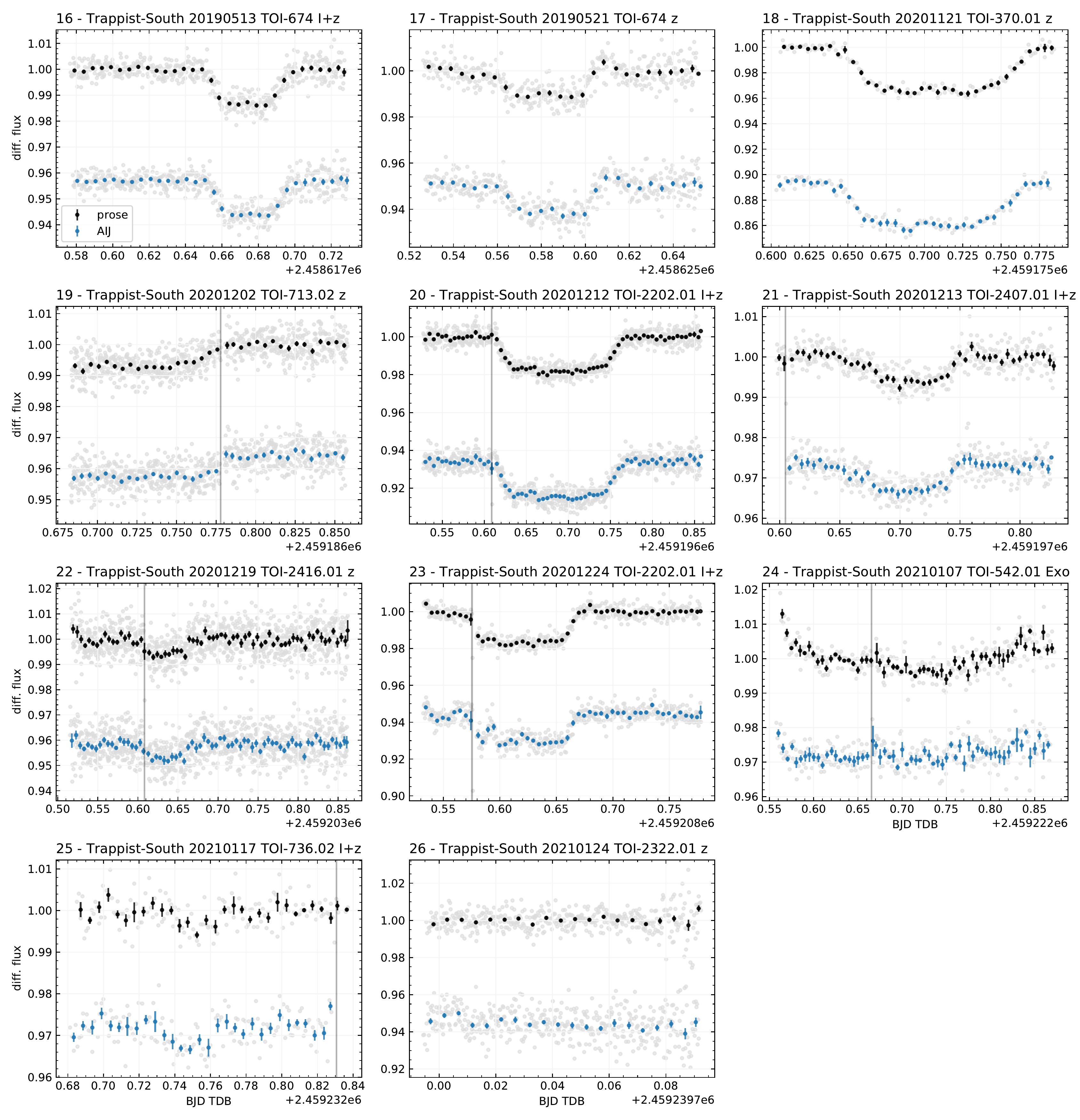}
  \caption{Comparison dataset light curves obtained from prose (black) and \AIJ{} (blue). Differential fluxes are plotted in light gray and supplemented by binned colored points (binning of 0.05 days) with an error bar that corresponds to the standard deviation of points within each bin. When it is relevant, the vertical gray line indicates the time of meridian flip. Light curves shown in this figure are detrended from a meridian flip step signal and an order two polynomial of time, simultaneously fitted with a transit signal as described in \autoref{metrics}. We bring the reader's attention to the fact that, unlike the signals shown in this figure, the four metrics presented in \autoref{metrics} are evaluated on light curves cleaned from transit signals. Observations indexes (as defined in \autoref{tab:aij-dataset}) are indicated in the title of each plot)}
  \label{fig:all_lcs}
\end{figure}

\autoref{tab:aij-dataset} summarizes our comparison dataset. \autoref{fig:all_lcs} shows the light curves obtained from these observations, using \prose{} and \AIJ{}.

\begin{table}[H]
  \centering
    \bgroup
    \footnotesize
    \def\arraystretch{1.3}
    \begin{tabular}{|clcccc|}
      \hline
      index & telescope & date & target & filter & type\\
      \hline
      1 & TRAPPIST-South & 2018 10 01 & TOI-145 & z & other \\
      2 & TRAPPIST-South & 2018 10 18 & TOI-142 & I+z & transit \\
      3 & TRAPPIST-South & 2018 11 03 & TOI-169 & B & transit \\
      4 & TRAPPIST-South & 2018 11 25 & TOI-207 & I+z & transit \\
      5 & TRAPPIST-South & 2018 11 28 & TOI-212 & z & transit \\
      6 & TRAPPIST-South & 2018 11 29 & TOI-171 & B & transit \\
      7 & TRAPPIST-South & 2018 12 14 & TOI-273 & I+z & other \\
      8 & TRAPPIST-South & 2018 12 15 & TOI-270 & z & transit \\
      9 & TRAPPIST-South & 2018 12 19 & TOI-150 & B & transit \\
      10 & TRAPPIST-South & 2018 12 27 & TOI-270 & z & transit \\
      11 & TRAPPIST-South & 2019 01 12 & TOI-350 & I+z & transit \\
      12 & TRAPPIST-South & 2019 02 16 & TOI-442 & z & transit \\
      13 & TRAPPIST-South & 2019 03 24 & TOI-500 & B & transit \\
      14 & TRAPPIST-South & 2019 03 27 & TOI-504 & RC & transit \\
      15 & TRAPPIST-South & 2019 04 17 & TOI-602 & z & transit \\
      16 & TRAPPIST-South & 2019 05 13 & TOI-674 & I+z & transit \\
      17 & TRAPPIST-South & 2019 05 21 & TOI-674 & z & transit \\
      18 & TRAPPIST-South & 2020 11 21 & TOI-370.01 & z & transit \\
      19 & TRAPPIST-South & 2020 12 02 & TOI-713.02 & z & transit \\
      20 & TRAPPIST-South & 2020 12 12 & TOI-2202.01 & I+z & transit \\
      21 & TRAPPIST-South & 2020 12 13 & TOI-2407.01 & I+z & transit \\
      22 & TRAPPIST-South & 2020 12 19 & TOI-2416.01 & z & transit \\
      23 & TRAPPIST-South & 2020 12 24 & TOI-2202.01 & I+z & transit \\
      24 & TRAPPIST-South & 2021 01 07 & TOI-542.01 & Exo & other \\
      25 & TRAPPIST-South & 2021 01 17 & TOI-736.02 & I+z & transit \\
      26 & TRAPPIST-South & 2021 01 24 & TOI-2322.01 & z & other \\     
      \hline
    \end{tabular}
    \vspace{0.3cm}
    \caption{Observations used for comparison with \AIJ{}.}
    \label{tab:aij-dataset}
    \egroup
  \end{table} 

\begin{figure}[H]
  \hspace{-1.4cm}
  \includegraphics[width=1.1\linewidth]{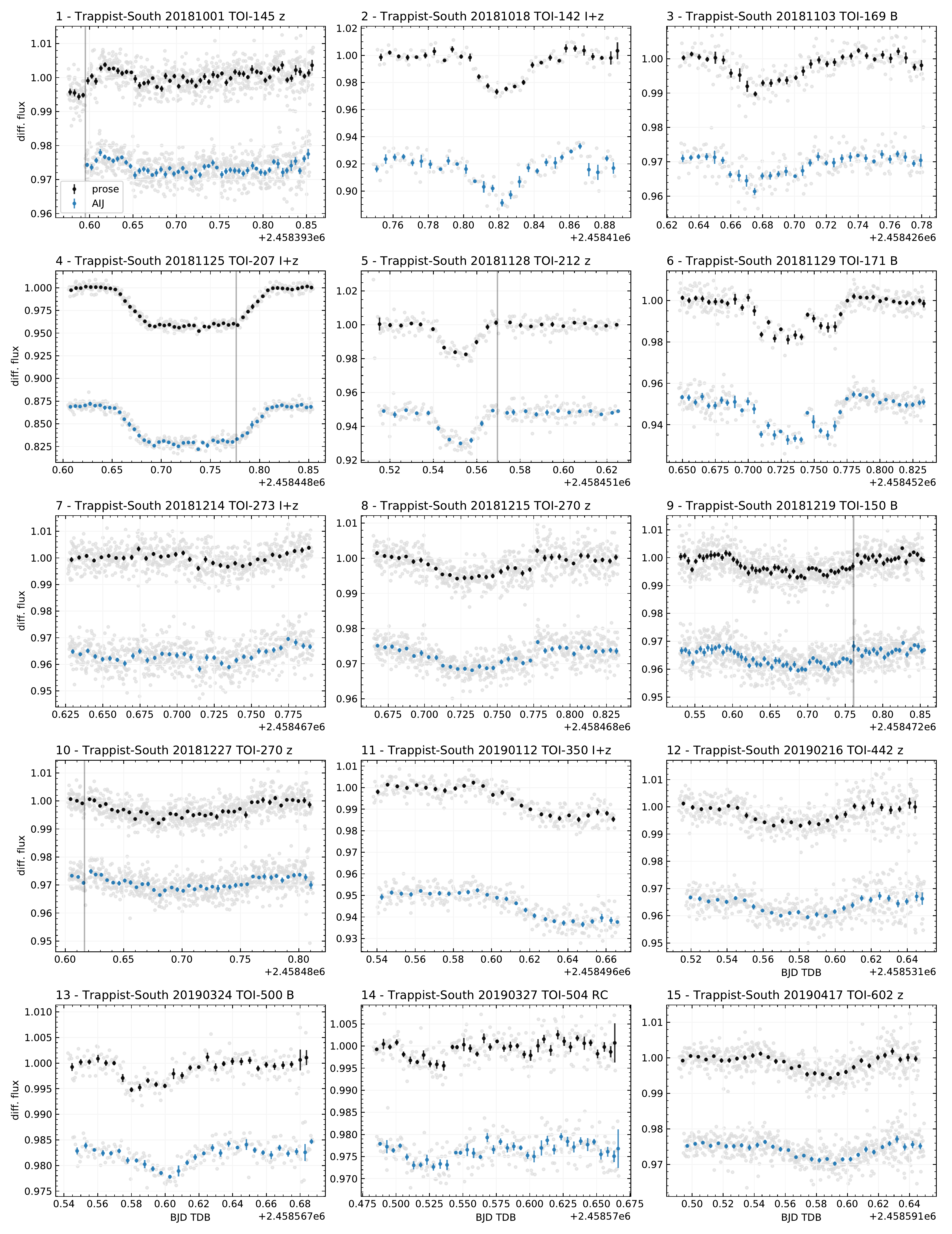}
\end{figure}
\newpage

\newpage
\section{TESS reports} \label{apreports}

\begin{figure}[H]
    \hspace{-2.5cm}
		\includegraphics[width=1.3\linewidth]{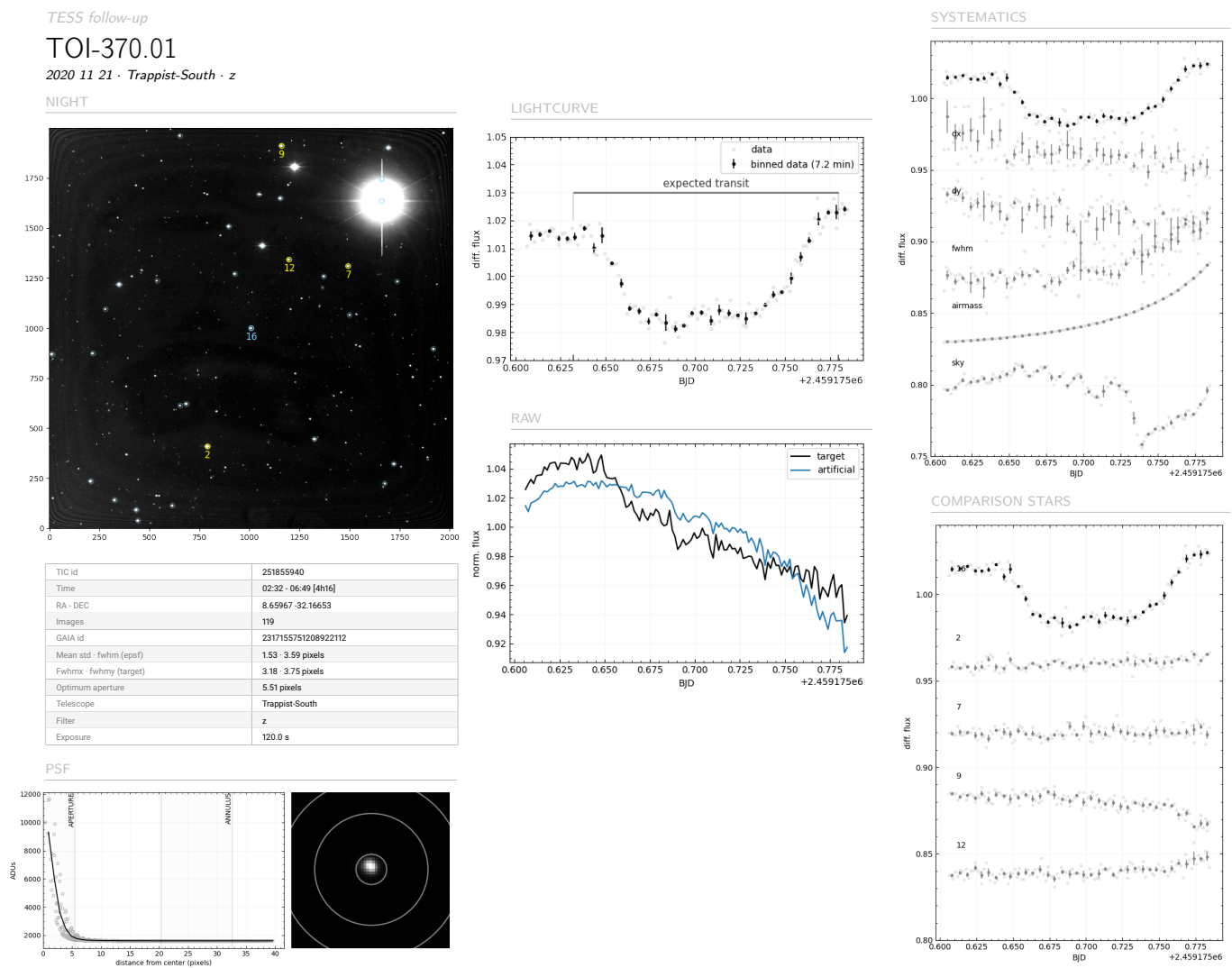}
		\caption{\prose{} reports start with a \textit{Summary} page, which intends to provide a quick look to an observation and its main products. The left side of this page features a stack image with the detected stars, among which are highlighted the target star (in blue) and comparison stars (yellow) used to build the differential light curves. At the bottom left, a cutout around the target as well as a radial PSF is plotted and the corresponding aperture overlaid. The rest of the page displays the raw and differential flux of the target star, as well as external parameters time-series (e.g. airmass) and comparison stars' light curves.}
		\label{fig:report1}
	\end{figure}

\begin{figure}[H]
    \hspace{-1.5cm}
		\includegraphics[width=1.4\linewidth, page=2]{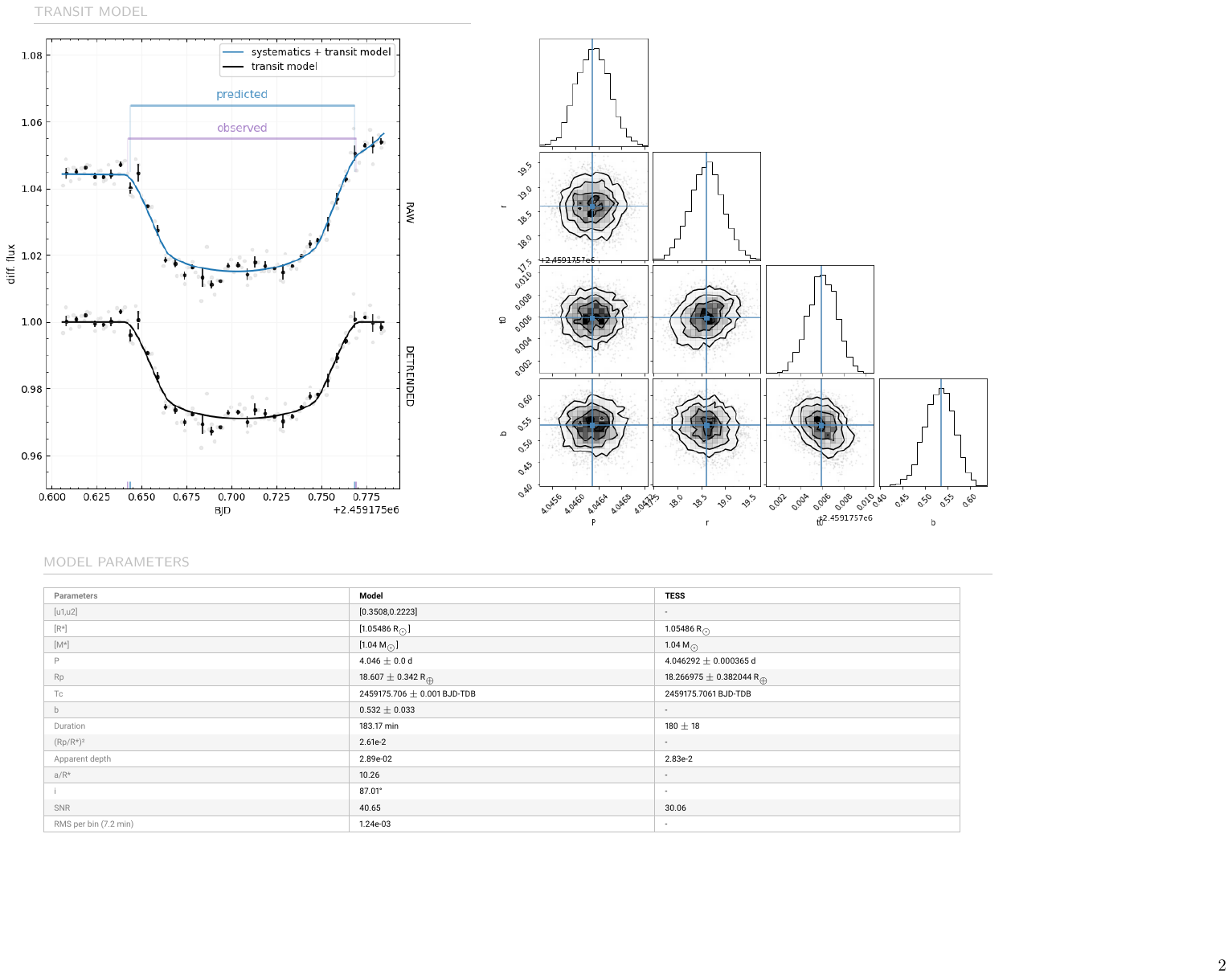}
  \caption{The \textit{Transit model} report page shows a model of the transit light curve together with the inferred transit parameters (here a corner plot has been added manually). This model was inferred using the \texttt{exoplanet} Python package \citep{exoplanet} and the corner plot done with \texttt{corner} \citep{corner}.}
  \label{fig:report3}
\end{figure}

\autoref{fig:report1} and \autoref{fig:report3} show elements of a TESS follow-up report produced with \prose{} using \LaTeX. While the first \textit{Summary} page is automatically generated, other pages are pre-filled and can be freely complemented with figures and text.

\newpage

\end{document}